\documentclass[graybox, natbib]{svmult}
\bibpunct{(}{)}{;}{a}{}{,} 

\usepackage{amsmath}

\usepackage{mathptmx}       
\usepackage{helvet}         
\usepackage{courier}        
\usepackage{type1cm}        
\usepackage{amsmath}         

\usepackage{makeidx}         
\usepackage{graphicx}        
\usepackage{multicol}        
\usepackage[bottom]{footmisc}
\usepackage[normalem]{ulem}	
\usepackage{hyperref}  
\usepackage{soul}   

\def\beq{\begin{equation}}
\def\eeq{\end{equation}}

\def\bs{\begin{split}}
\def\es{\end{split}}

\def\bea{\begin{eqnarray}}
\def\eea{\end{eqnarray}}

\makeindex             

\begin{document}

\title*{The fundamentals of quantum machine learning}
\author{Bing Huang, Nadine O. Symonds, O. Anatole von Lilienfeld}
\institute{Institute of Physical Chemistry and National Center for Computational Design and Discovery of Novel Materials (MARVEL),
Department of Chemistry, University of Basel, Klingelbergstrasse 80, 4056 Basel, Switzerland, \email{anatole.vonlilienfeld@unibas.ch}}
%
%
\maketitle
\abstract{Within the past few years, we have witnessed the rising of quantum machine learning (QML) models which infer electronic properties of molecules and materials, rather than solving approximations to the electronic Schr\"odinger equation.
The increasing availability of large quantum mechanics reference data sets have enabled these developments.  We review the basic theories and key ingredients of popular QML models such as choice of regressor, data of varying trustworthiness, the role of the representation, and the effect of training set selection. Throughout we emphasize the indispensable role of learning curves when it comes to the comparative assessment of different QML models.}

\section{Introduction}

\textcolor{black}{Society is becoming increasingly aware of its desperate need for new molecules and materials, be it new antibiotics, or efficient energy storage and conversion materials. Unfortunately, chemical compounds reside in, or rather hide among, an unfathomably huge number of possibilities, also known as chemical compound space (CCS). CCS is the set of stable compounds which can be obtained through all combinations of chemical elements and interatomic distances.
For medium-sized drug-like molecules CCS is believed to exceed 10$^{60}$~\citep{CCS}. Exploration in CCS and locating the ``optimal'' compounds is thus an extremely difficult, if not impossible, task. Typically, one needs to constrain the search domain in CCS and obtains certain pertinent properties of compounds within the subspace, and then choose the compounds with properties which come closest to some preset criteria as potential candidates for subsequent updating or validation. Of course, one can conduct experiments for each compound. Alternatively, one can also attempt to estimate its properties using modern atomistic simulation tools which, within one approximation or the other, attempt to solve Schr\"odinger's equation on a modern powerful computer.
}

\textcolor{black}{The latter approach is practically more favorable and referred as high-throughput (HT) computational screening~\citep{greeley2006}. In spite of its popularity, it is inherently limited by the computational power accessible considering that 1) the number of possible compounds is much larger than what HT typically is capable of dealing with ($\sim$$10^3$) and 2) often very time-consuming explicitly electron correlated methods are necessary to reach chemical accuracy (1 kcal/mol for energies), with computational cost often scaling as $O(N^6)$ ($N$ being the number of electrons, a measure of the system size). Computationally more efficient methods generally suffer from rather weak predictive power. They range from force-fields and semi-empirical molecular orbital methods, density functional theory (DFT) methods to so-called linear scaling methods which assume locality by virtue of fragments or localized orbitals~\citep{FMO}. It remains an outstanding challenge within conventional computational chemistry that efficiency and accuracy apparently cannot coexist. }

\textcolor{black}{To tackle this issue,  Rupp, et al~\citep{CM} introduced a machine learning (ML) Ansatz in 2012, capable of predicting atomization energies of out-of-sample molecules fast and accurately for the the first time. By now many subsequent studies showed that ML models enable fast and yet arbitrarily accurate prediction for \emph{any} quantum mechanical property. This is no ``free lunch'', however, the price to pay consists of the acquisition of a set of pre-calculated training data sets which must be sufficiently representative and dense. }

So what is machine learning? 
\textcolor{black}{It is a field of computer science that gives computers the ability to learn without being explicitly programmed. ~\citep{WhatIsML}} \textcolor{black}{
Among the broad categories of ML tasks, we focus on a type called supervised learning with continuous output, which infers a function from labeled training data. Putting it formally, given a set of $N$ training examples of the form \{($x_1$,$y_1$), ($x_2$,$y_2$), $\cdots$, ($x_N$,$y_N$)\} with $x_i$ and $y_i$ being respectively the input (the representation) and output (the label) of example $i$, a ML algorithm models the implicit function $f$ which maps input space $X$ to label space $Y$. The trained model can then be applied to predict $y$ for a new input $x$ (belonging to the so-called test set) absent in the training examples.}
\textcolor{black}{For quantum chemistry problems, the input of QML (also called representation) is usually a vector/matrix/tensor directly obtained from composition and geometry \{$Z_I$, $\mathbf{R}_I$\} of the compound; while the label could be \emph{any} electronic property of the system, notably the energy. The function $f$ is implicitly encoded in terms of the non-relativistic Schr\"odinger equation (SE) within the Born-Oppenheimer approximation, $\hat{H}\Psi = E\Psi$, whose exact solution is unavailable for all but the smallest and simplest systems. To generate training data, methods with varied degrees of approximation have to be used instead, such as the aforementioned DFT, QMC, etc.}
 

\textcolor{black}{Given a specific pair of $X$ and $Y$, there are multiple strategies to learn the implicit function $f:X\to Y$. Some of the most popular ones are artificial neural network (ANN, including its various derivatives, such as convolutional neural network) and kernel ridge regression (KRR, or more generally Gaussian process regression). }
\textcolor{black}{ Based on a recent benchmark paper \citep{googlePaper2017}, KRR and ANN are competitive in terms of performance.  KRR, however, has the great advantage of simplicity in interpretation and ease in training, provided an efficient representation is used. Within this chapter, we therefore focus on KRR or Gaussian processes exculsively. See section 2 for more details. }

\textcolor{black}{Often, each training example is represented by a pair ($x_i,y_i$). However, multiple $\{y_{j}\}_i$ can also be used, e.g.~when multiple labels are available for the same molecule, possibly resulting from different levels of theory. The latter situation can be very useful for obtaining highly accurate QML models with scarcely available accurate training data and coarse data being easy to obtain. Multi-fidelity methods  take care of such cases and will be discussed in section 3.
}

\textcolor{black}{Once the suitable QML model is selected, be it either in terms of ANN, KRR or in terms of a multi-fidelity approach, two additional key factors will have a strong impact on the performance: The materials representation and the selection procedure of the training set. 
The representation of any compound should essentially result from a bijective map which uses as input the same information which is also used in the electronic Hamiltonian of the system, i.e.~compositional and structural information \{$Z_I$,$\mathbf{R}_I$\} as well as electron number. The representation is then typically formatted into a vector which can easily be processed by the computer. Some characteristic representations, introduced in the literature, are described in section 4, where we will see how the performance of QML models can be enhanced dramatically by accounting for more of the underlying physics. 
In section 5, further improvements in QML performance are discussed resulting from 
rational training set selection, rather than from random sampling. }

\textcolor{black}{Having introduced the basics of ML, we are motivated to point out two aspects of ML that may not be obvious for better interpretation of how ML works: 1) ML is an inductive approach based on rigorous implementation of inductive reasoning and it does not require \emph{any} a priori knowledge about the aforementioned implicit function $f$ (see section 2), though some insight of what $f$ may look like is invaluable for rational design of representation  (see section 4); 2) ML is of interpolative nature, that is, to make reasonable prediction, the new input must fall into the interpolating regime. Furthermore, as more training examples are added to the interpolating regime, the performance of the ML model can be systematically improved for a quantified representation (see section 4). 
}

As a sidenote, we would like to mention the importance of turning basic theories of QML into user-friendly and efficient code, 
so that anybody in the community can benefit from these new developments. 
Among multiple options, the recently released QML code~\citep{QMLcode1} covers a substantial number of QML models, 
some of which are presented in the following sections.

\section{Gaussian process regression}
In this section, we discuss the basic idea of data driven prediction of labels: the Gaussian process regression (GPR). In the case of a global representation (i.e., the representation of any compound as a single vector, see section 4 for more details), the corresponding QML model takes the same form as in kernel ridge regression (KRR), also termed the global model. GPR is more general than KRR in the sense that GPR is equally applicable to local representations (i.e., the representation of any compound as a 2D array, with each atom in its environment represented by a single vector, see section 4 for more details). Local GPR models can still successfully be applied when it comes to the prediction of extensive properties (e.g., total energy, isotropic polarizability, etc.) which profit from near-sightedness. The locality can be exploited for the generation of scalable GPR based QML models which can be used to estimate extensive properties of very large systems.

\subsection{The global model} \label{sec:GPR}
Here we review the Bayesian analysis of the nonlinear regression model~\citep{GP} 
with Gaussian noise $\epsilon$:
\beq 
\mathbf{y} = \phi(\mathbf{x})^{\top}\mathbf{w} + \varepsilon,
\eeq 
where $\mathbf{x}\in \mathbf{X}$ is the representation, $\mathbf{w}$ is a vector of weights,
and $\phi(\mathbf{x})$ is the basis function (or kernel) which maps a $D$-dimensional input vector $\mathbf{x}$ into an $N$ dimensional feature space. This is the space into which the input vector is mapped, e.g., for an input vector $\mathbf{x}_1 = (x_{11},x_{12})$ with $D=2$, its feature space could be $\phi(\mathbf{x}_1)=(x_{11}^2,x_{11}x_{22},x_{22}x_{11},x_{22}^2)$ with $N=4$. $\mathbf{y}$ is the label, i.e.~the observed property of target compounds. 
We further assume that the noise $\epsilon$ follows an independent, identically distributed (iid) Gaussian distribution with zero mean and variance $\lambda$, i.e., $\epsilon \sim \mathcal{N}(0, \lambda)$, 
which gives rise to the probability density of the observations given the parameters $\mathbf{w}$, or the likelihood
\beq \label{eq:likelyhood}
p(\mathbf{y}|\mathbf{X},\mathbf{w}) = \prod_{i=1}^n \mathcal{N}(\phi(\mathbf{x}_i)^{\top}\mathbf{w}, \lambda I) = \mathcal{N}(\phi(\mathbf{X})^{\top}\mathbf{w}, \lambda I),
\eeq 
where $\phi(X)$ is the aggregation of columns $\phi(\mathbf{x})$ for all cases in the training set. Now we put a zero mean Gaussian prior with covariance matrix $\Sigma_p$ over $\mathbf{w}$ to express our beliefs about the parameters before we look at the observations, i.e., $\mathbf{w} \sim \mathcal{N}(0, \mathrm{\Sigma}_p)$. Together with Bayes' rule
\beq
p(\mathbf{w}|\mathbf{y},\mathbf{X}) = \frac{p(\mathbf{y}|\mathbf{X},\mathbf{w})p(\mathbf{w})}{p(\mathbf{y}|\mathbf{X})}
\eeq
\beq \label{eq:Bayes}
p(\mathbf{y}|\mathbf{X}) = \int p(\mathbf{y}|\mathbf{X},\mathbf{w})p(\mathbf{w}) d\mathbf{w},
\eeq 
distribution of $\mathbf{w}$ can be updated as
\beq \label{eq:Posterior}
p(\mathbf{w}|\mathbf{X},\mathbf{y}) \sim \mathcal{N}(\bar{\mathbf{w}}=\lambda^{-1}A^{-1}\phi(\mathbf{X})\mathbf{y}, A^{-1}) 
\eeq 
where $A=\lambda^{-1}\phi(\mathbf{X})\phi(\mathbf{X})^{\top} + \mathrm{\Sigma}_p^{-1}$. The updated $\mathbf{w}$ is called the posterior with mean $\bar{\mathbf{w}}$. Thus, similar to equation (\ref{eq:Bayes}), the predictive distribution for $\mathbf{y}_* = f(\mathbf{x_*})$ is
\beq \label{eq:Pred0}
p(\mathbf{y}_*|\mathbf{x}_*,\mathbf{X},\mathbf{y}) = \int p(\mathbf{y}_*|\mathbf{x}_*,\mathbf{w})p(\mathbf{w}|X,\mathbf{y}) d\mathbf{w}.
\eeq 
Substituting equation (\ref{eq:likelyhood}) and (\ref{eq:Posterior}) into equation (\ref{eq:Pred0}),
\beq 
p(\mathbf{y}_{*}|\mathbf{x}_*, \mathbf{X},\mathbf{y}) = \mathcal{N}(\lambda^{-1}\phi(\mathbf{x}_*)^{\top} A^{-1}\phi(\mathbf{X})\mathbf{y}, \phi(\mathbf{x}_*)^{\top} A^{-1}\phi(\mathbf{x}_*)),
\eeq 
which can be further simplified to $p(\mathbf{y}_{*}|\mathbf{x}_*, \mathbf{X},\mathbf{y}) = \mathcal{N}(\bar{\mathbf{y}}_*, \bar{\lambda}$) with $\bar{\mathbf{y}}_*$ and $\bar{\lambda}$ being respectively 
\beq \label{eq:GP_global}
\bar{\mathbf{y}}_* = K(\mathbf{x}_*,\mathbf{X}) (K(\mathbf{X},\mathbf{X}) + \lambda I)^{-1} \mathbf{y},
\eeq
\beq
\bar{\lambda} = K(\mathbf{x}_*,\mathbf{x}_*) - K(\mathbf{x_*},\mathbf{X})(K(\mathbf{X},\mathbf{X})+\lambda I)^{-1}K(\mathbf{X},\mathbf{x_*}),
\eeq 
where $I$ is the identity matrix, $K(\mathbf{X},\mathbf{X}) = \phi '(\mathbf{X})^{\top}\phi '(\mathbf{X})$ ($\phi '(\mathbf{X}) =  \mathrm{\Sigma}_p^{1/2} \phi(\mathbf{X})$) is the kernel matrix (also called covariance matrix, abbreviated as Cov). It's not necessary to know $\phi$ explicitly, their existence is sufficient. Given a Gaussian basis function, i.e., $\phi '(x)=\exp (-(x-x_0)^2/(2l^2))$ with $x_0$ and $l$ being some fixed parameters, it can be easily shown that the $(i,j)$-th element of kernel matrix $K$ is 
\beq \label{eq:kernel}
k(\mathbf{x}_i, \mathbf{x}_j) = \exp \left (-\frac{1}{2}\frac{||\mathbf{x}_i - \mathbf{x}_j||_2^2}{\sigma^2} \right ),
\eeq 
where $||\cdot||_p$ is the $L_p$ norm, $\sigma$ is the kernel width determining the characteristic length scale of the problem. Note that we have avoided the infeasible computation of feature vectors of infinite size by using some kernel function $k$. This is also called the kernel trick. Other kernels can be used just as well, e.g.~the Laplacian kernel, 
$k(\mathbf{x}_i, \mathbf{x}_j) = \exp \left (- \frac{||\mathbf{x}_i - \mathbf{x}_j||_1}{\sigma} \right )
$.

Rewriting equation \ref{eq:GP_global}, we arrive at a more concise expression
in matrix form,
\beq \label{eq:gk}
\mathbf{y}_* = K(\mathbf{X}_*,\mathbf{X})\mathbf{c},
\eeq
where $\mathbf{c}$ is the regression coefficient vector,
\beq \label{eq:coeffs}
\mathbf{c} = (K(\mathbf{X},\mathbf{X}) + \lambda I)^{-1}\mathbf{y}.
\eeq

Equation (\ref{eq:gk}) can also be obtained by minimizing the cost function $C(\mathbf{w}) = \frac{1}{2}\sum_i (y_i - \mathbf{w}^{\top}\phi(\mathbf{x}_i))^2 + \frac{\lambda}{2} ||\mathbf{w}||_2^2$,
with respect to $\mathbf{w}$. 
Note that $L_2$ regularization is used here, together with a regularization parameter $\lambda$ acting as a weight to balance minimizing the sum of squared error (SSE) and limiting the complexity of the model. This eventually leads to a model called kernel ridge regression (KRR) model. 

All variants of these global models, however, suffer from the scalability problem for extensive properties of the system such as energy, i.e., the prediction error grows systematically with respect to query system size (predicted estimates will tend towards the mean of the training data while extensive properties grow). This limitation is due to the interpolative nature of global ML models, that is, the predicted query systems and their properties must lie within the domain of training data. 

\subsection{The local version} \label{sec:GPR_local}
The scalability problem can be overcome by working with local, e.g.~atomic, representations. This relies on the idea that one can decompose a global extensive property of the system into local contributions. Among the many ways to partition systems into building-blocks, we select the atom-in-molecule (AIM) idea, put forth many years ago by Bader~\citep{AIM}. For the total energy ($E$) of the system, it is usually expressed as a sum over atomic energies ($e$), 
\beq \label{eq:aim}
E = \sum_I e^I = \sum_I \int_{\Omega_{I}} \langle \Psi | \hat{H } | \Psi \rangle d^3r
\eeq 
where $\Omega_I$ is the atomic basin determined by the zero-flux condition of the electron density,
\beq
\nabla \rho(\mathbf{r_s})\cdot \mathbf{n}(\mathbf{r_s}) = 0,~~\mathrm{for~every~point~}\mathbf{r_s}~\mathrm{on~the~surface}~\mathrm{S}(\mathbf{r}_\mathrm{s})
\eeq 
where $\mathbf{n}(\mathbf{r_s})$ is the unit vector normal to the surface at $\mathbf{r}_\mathrm{s}$. The advantage of using Bader's scheme is that the total energy is exactly recovered, and that, at least in principle, it includes all short- and long-ranged bonding, i.e.~covalent as well as non-covalent (e.g., van der Waals interaction, Coulomb interaction, etc.). Furthermore, due to nearsightedness of atoms in electronic system~\citep{nearsightedness}, atoms with similar local chemical environments contribute a similar amount of energy to the total energy. Using the notion of alchemical derivatives, this effect, a.k.a.~chemical transferability, has recently been demonstrated numerically~\cite{StijnPNAS2017}.
Thus it is possible to learn effective atomic energies based on a representation of the local atoms. Unfortunately, the explicit calculation of local atoms is computationally involved (the location of the zero-flux plane is challenging for large molecules), making this approach less favorable. 
Instead, we can also assume that the aforementioned Bayesian model is applicable to atomic energies as well, i.e.,
\beq \label{eq:bayesian_atom}
e^I = \phi(\mathbf{x}^I)^{\top}\mathbf{w} + \varepsilon 
\eeq 
where $\mathbf{x}^I$ is an atomic representation of atom $I$ in a molecule.
By summing up terms on both sides in equation~\ref{eq:bayesian_atom}, we have 
\bea
E = \sum_I \phi(\mathbf{x}^I)^{\top}\mathbf{w} + \varepsilon.
\eea
Following Bartok~\citep{GAP}, the covariance of the total energies of two compounds can be expressed as
\beq \label{eq:lk}
K_{ij} = \mathrm{Cov}(E_i, E_j) = \mathrm{Cov}(\sum_I e_i^I, \sum_J e_j^J) = \sum_I \sum_J \mathrm{Cov}(e_i^I,e_j^J) = \sum_I \sum_J k(\mathbf{x}_i^I, \mathbf{x}_j^J)
\eeq
where $I$ and $J$ run over all the respective atomic indices in molecule $i$ and $j$, and where $\mathbf{x}^I_i$ is the representation of atom $I$ in molecule $i$. 

By inserting equation (\ref{eq:lk}) in equation (\ref{eq:gk}), we arrive at the formula for the energy prediction of a molecule $*$ out-of-sample,
\beq
E_* = \sum_{i} c_i \sum_{I\in i} \sum_{J\in *} k(\mathbf{x}_i^I, \mathbf{x}_*^J)
\eeq
where $c_i = \sum_{j} ([K + \lambda I]^{-1})_{ij} E_j$. This equation can be rearranged,
\bea
E_* & = &\sum_{J\in *} \sum_{i} c_i \sum_{I\in i} k(\mathbf{x}_i^I, \mathbf{x}_*^J)
\; = \; \sum_{J\in *} e_*^J,
\eea
where the atomic contribution of atom $J$ to the total energy can be decomposed into a linear combination of contributions from each training compound $i$, weighted by its regression coefficient, 
\bea \label{eq:ae_krr}
e_*^J & = & \sum_i c_i \tilde{e}_{*_i}^J. 
\eea 
The ``basis-function'' $\tilde{e}_{*_i}^J$ in this expansion simply consist of the sum over kernel similarities between atom $J$ and atoms $I \in i$, where the contribution of atom $I$ grows with its similarity to atom $J$, 
\bea 
\tilde{e}_{*_i}^J  & = & \sum_{I} k(\mathbf{x}_i^I, \mathbf{x}_*^J).
\eea

We note in passing that the value of the covariance matrix element (i.e., equation (\ref{eq:lk})) increases when the size of either system $i$ or $j$ grows, indicating that the scalability issue can be effectively resolved.

\subsection{Hyper-parameters}

\textcolor{black}{Within the framework of GPR or KRR, there are two sets of parameters: 1) parameters that are determined via training, i.e., the coefficients $\mathbf{c}$ (see equation (\ref{eq:coeffs})), whose number grows with the training data; 2) hyperparameter whose value is set before the learning process begins, i.e., the kernel width $\sigma$ in equation (\ref{eq:gk}) and $\lambda$ in equation (\ref{eq:likelyhood}). 
}

\textcolor{black}{
As defined in section \ref{sec:GPR}, $\lambda$ measures the level of noise in the training data in GPR. Thus, if the training data is noise free, $\lambda$ can be safely set to zero or a value extremely close to zero (e.g., $1\times 10^{-10}$) to reach optimal performance. This is generally true for datasets obtained by typical quantum chemical calculations and the resulting training error is (almost) zero. Whenever there is noise in the data (e.g., from experimental measurements), the best $\lambda$ corresponds to some finite value depending on the noise level. The same holds for the training error. 
In terms of KRR, $\lambda$ seems to have a completely different meaning at first glance: the regularization parameter determining the complexity of the model. In essence, they amount to the same, i.e., a minute or zero $\lambda$ corresponds to the perfectly interpolating model which connects every single point in the training data, thus representing the most faithful model for the specific problem at hand. One potential risk is poor generalization to new input data (test data), as there could be ``overfitting'' scenarios for training sets. A finite $\lambda$ assumes some noise in the training data and the model can only account for this in an averaged way, thus the model complexity is simplified to some extend by lowering the magnitude of parameters $\mathbf{w}$ so as to minize the cost function $C(\mathbf{w})$. Meanwhile, some finite training error is introduced. To recap, the balance between SSE and regularization is vital, and reflected by a proper choice of $\lambda$.}

\textcolor{black}{Unlike $\lambda$, the optimal value of $\sigma$ ($\sigma_{\mathrm{opt}}$) is more dataset specific. Roughly speaking, it is a measure of the diversity of the dataset and controls the similarity (covariance matrix element) of two systems. Typically $\sigma_{\mathrm{opt}}$ gets larger when the training data expands into a larger domain. The meaning of $\sigma$ can be elaborated by considering two extremes: 1) when $\sigma$ approaches zero, the training data will be reproduced exactly, i.e., $c_i = y_i$, with high error for test data, i.e.~with deviation to mean; 2) when $\sigma$ is infinity, all kernel matrix elements will tend towards one, i.e., a singular matrix, resulting in large errors in both training and test. Thus, the optimal $\sigma$ can be interpreted as a coordinate scaling factor to render the kernel matrix well-conditioned. For example, ~\citep{one_kernel_Raghu} selected the lower bound of the kernel matrix elements to be 0.5. For a Gaussian kernel, this implies that $K_{\min} = \exp(-D_{\max} ^2/2\sigma_{opt}^2) \approx 0.5$, or $\sigma_{opt} \approx D_{\max}/\sqrt{2\ln 2}$, where $D_{\max}$ is the largest distance matrix element of the training data. Following the same reasoning, $\sigma_{opt} $ can be set to $D_{\max}/\ln 2$ for a Laplacian kernel. }

\textcolor{black}{The above heuristics are very helpful to quickly identify reasonable initial guesses for hyper-parameters for a new data set. Subsequently, the optimal values of the hyper-parameters should be fine-tuned through $k$-fold cross-validation (CV). The idea is to first split the training set into $k$ smaller sets and 1) for each of the $k$ subsets, a model is trained using the remaining $k-1$ subsets as training data;
the resulting model is tested on the remaining part of the data to calculate the predictive error); this step yields $k$ predictions, one for each fold.
2) The overall error reported by $k$-fold cross-validation is then the average of the above $k$ values. The optimal parameters will correspond to the ones minimizing the overall error.
This approach can become computationally demanding when $k$ and the training set size are large. But it is of major advantage in problem such as inverse inference where the number of samples is very small, and its systematic applications minimizes the likelihood of statistical artefacts.}

\subsection{Learning curves} \label{sec:LC}
To assess the predictive performance of a ML model, we need to know not only the prediction error ($\epsilon$, which can be characterized by the mean absolute error (MAE) or root mean squared error (RMSE) of prediction) for a specific training set, but also predictive errors for varied sizes of training sets. Therefore, we can monitor how much progress we have achieved after some incremental changes to the training set size ($N$) so as to extrapolate to see how much more training data is needed to reach a desirable accuracy. The plot of $\epsilon$ versus $N$ relationship is called the learning curve (LC), and  examples are shown in Fig. 1. 
(note that only test error, i.e., MAE for the prediction of new data in test set, is shown; training errors are always zero or minute for noise-free training data). 

\begin{figure} \label{fig:LC}
  \centering
  \includegraphics[width=0.5\columnwidth]{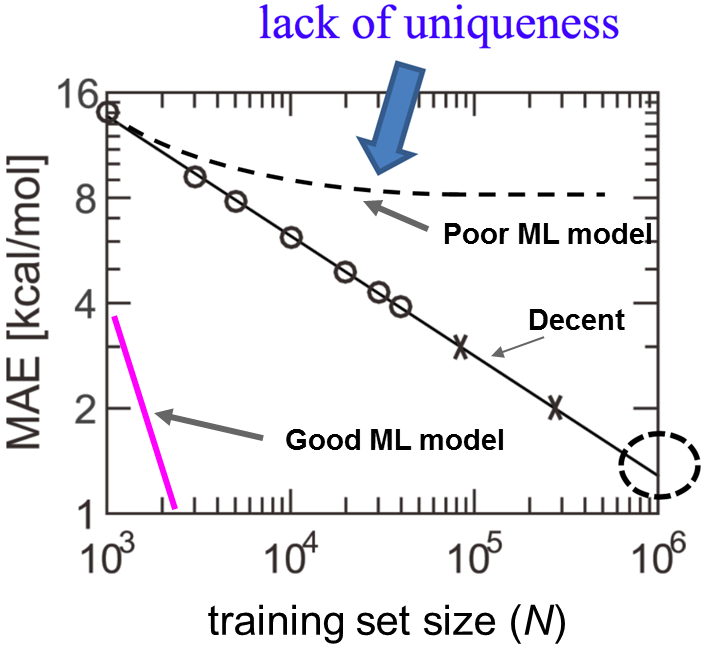}
  \caption{Three representative learning curves with distinguished relative performance.}
\end{figure}


Multiple factors control the shape of learning curve, one of which is the choice of representation. If the representation cannot uniquely encode the molecule, i.e., there may exist cases that two different molecules share the same input vector $\mathbf{x}_i$ but with different molecular properties, then it causes ambiguity to the ML algorithm (see more details in section \ref{sec:GoodRepr}) and may consequently lead to no learning at all, as illustrated by the dashed curve in Fig. 1, 
with distinguishable flattening out behavior at larger training set sizes, resulting in poor ML performance.

In the case of a unique representation, according to~\citep{fasshauer2016kernel}, it can be proved
that for kernel based approximation, when the training set size $N$ is sufficiently large, 
the predictive error is proportional to the so-called ``fill distance'' or \textit{mesh norm} $h_\mathbf{X}$, defined as
\beq
h_\mathbf{X} = \underset{\mathbf{x}\in \mathrm{\Omega} }{\sup }~\underset{\mathbf{x}_j \in \mathbf{X}}{\min}~||\mathbf{x}-\mathbf{x}_j||_2
\eeq 
where ``$\sup$" stands for the supremum (or the least upper bound) of a subset, $\mathbf{x}$ is again the representation of any training instance as an element of the training set $\mathbf{X}$, $\Omega$ represents the domain of studied systems (i.e., potential energy surface domain for chemistry problems). Clearly from the definition, fill distance describes the geometric relation of the set $\mathbf{X}$ to the domain $\mathrm{\Omega}$ and quantifies how densely $\mathbf{X}$ covers $\mathrm{\Omega}$. Furthermore, fill distance intrinsically contains a dimension dependence $d$, that is, $h_\mathbf{X}$ scales roughly as $N^{-1/d}$ if $\mathbf{x}$ are uniform or random grid points in a $d$ dimensional space.

Apart from the exponent, there should also be a prefactor, thus the leading term of the overall predictive error can be described as $b * N^{-a/d}$, where $a$ in the exponent is a constant. Therefore, to visualize the error vs. $N$, a log-log scale is the most convenient for which the learning curve can be represented by a linear relationship: $\log (\epsilon) \approx \log (b) - \frac{a}{d} \log (N)$, thus $a/d$
quantifies the rate of learning, while the prefactor $\log(b)$ is the vertical offset of the learning curve. 
Through a series of numerical calculations of learning a 1D Gaussian function as well as ground state properties of molecules with steadily improving physics encoded in the representation, it has been found~\citep{baml} that the offset $\log(b)$ is a measure of  target property similarity, which is defined as the deviation of proposed model (corresponding to the representation used) from the true model~\citep{baml}. While in general, we do not know the true function (machine learning would be meaningless if we did) we often do have considerable knowledge about relative target similarity of different representations.

Applying the findings above to chemistry problems, we can thus obtain some insight in how learning curves will behave. Several observations can be explained: First, the learning rate would be almost a constant or changes very little when different unique representations are used, as the rate depends primarily on the domain spanned by molecules considered in the potential energy surface. Secondly, for a series of isomers
it is much easier to learn their properties in their relaxed equilibrium state than in a distorted geometry. 

The limitation that the learning rate will not change much for random sampling with unique representations seems to be an big obstacle towards more efficient ML predictions, meaning that developing better representation (to lower the offset) can become very difficult even if substantial effort has been invested. However, is it possible to break this curse, reaching an improved learning curve as illustrated by the pink line in Fig. 1? 
We believe that this should be possible. Note how the linear (log-log) learning curve is obtained for {\em statistical} models. This implies that there must be `redundancy' in the training data;
and if we were able to remove those redundancies {\em a priori}, we might very be able to boost the performance and observe superior LCs, such as the pink line in Fig. 1
with large learning rates. In such a case, statistics is unlikely to hold and the LC may be just a monotonically decreasing function, possibly also just a damped oscillator, rather than a line. Strategies for rational sampling will be elaborated in detail in  section 5.

\section{Multi-level learning}

\textcolor{black}{By default, we assume for each $\mathbf{x}_i \in \mathbf{X}$ there exists one corresponding $y_i \in Y$ in the training examples. It makes perfect sense if $Y$ is easy to compute, i.e., in the circumstance that a relatively low accuracy of $Y$ suffices (e.g., PBE with a medium sized basis set). It is also possible that a highly accurate reference data is required (e.g.,  CCSD(T) calculations with a large basis set) so as to achieve highly reliable predictions. Unfortunately, we can only afford few highly accurate ${\bf x}$ and $y$'s for training considering the great computational burden. In this situation one can take great advantage of the $y$'s with lower levels of accuracy which are much easier to obtain. Models which shine in this kind of scenario are called multi-fidelity, where reference data based on a high (low) level of theory is said to have high (low) fidelity. The nature of this approach is to explore and exploit the inherent correlation among data sets with different fidelities. Here we employ Gaussian process as introduced in section 2 to explain the main concepts and mathematical structure of multi-level learning.}

\subsection{Multi-fidelity}
For the sake of clarity and simplicity, we focus only on two levels of fidelity, the mathematical formulation stated below can be easily generalized to more fidelities. We consider two datasets with different level of fidelity: \{$\mathbf{X}, \mathbf{y}^{(1)}$\} (in which the pairs of data are $(\mathbf{x}_1, y_1^{(1)}), (\mathbf{x}_2, y_2^{(1)}),~\dots$) and \{$\mathbf{X}, \mathbf{y}^{(2)}$\}, where $\mathbf{y}^{(2)}$ has a higher level of fidelity. The number of data points in the two sets are respectively $N_1$ and $N_2$ and $N_1 > N_2$, reflecting the fact that high-fidelity data are scarce. We consider the following autoregressive model proposed by Kennedy and O'Hagan~\citep{kennedy2000}:
\beq 
\mathbf{y}^{(2)} = \rho \mathbf{y}^{(1)} + \delta ^{(2)}
\eeq 
where $\mathbf{y}^{(1)}$ and $\delta ^{(2)}$ are two independent Gaussian processes, i.e.,
\bea
\mathbf{y}^{(1)} &\sim& \mathcal{N}(0, K_1(\mathbf{X},\mathbf{X})) = \mathcal{N}(0,\mathrm{Cov}( \mathbf{y}^{(1)},\mathbf{y}^{(1)})) = \mathcal{N}(0, K_{1})\\ 
\delta ^{(2)} &\sim& \mathcal{N}(0, K_2(\mathbf{X}, \mathbf{X}) = \mathcal{N}(0, \mathrm{Cov}(\delta ^{(2)},\delta ^{(2)})) = \mathcal{N}(0, K_{2}).
\eea
That $\mathbf{y}^{(1)}$ and $\delta ^{(2)}$ are independent (notated as $\mathbf{y}^{(1)} \perp \delta ^{(2)}$) indicates that the mean of $\mathbf{y}^{(1)}\delta^{(2)}$ satisfies $\mathbf{E}[\mathbf{y}^{(1)}\delta^{(2)}] = \mathbf{E}[\mathbf{y}^{(1)}]\mathbf{E}[\delta^{(2)}]$
and thus the covariance between $\mathbf{y}^{(1)}$ and $\delta^{(2)}$ is zero, i.e., Cov($\mathbf{y}^{(1)}, \delta^{(2)}) = \mathbf{E}[\mathbf{y}^{(1)}\delta^{(2)}] - \mathbf{E}[\mathbf{y}^{(1)}]\mathbf{E}[\delta^{(2)}] = 0$.
Therefore, $\mathbf{y}^{(2)}$ is also a Gaussian process with mean 0 and covariance 
\bea
\mathrm{Cov}(\mathbf{y}^{(2)}, \mathbf{y}^{(2)}) &=& K_{22} = \mathrm{Cov}(\rho \mathbf{y}^{(1)} + \delta ^{(2)}, \rho \mathbf{y}^{(1)} + \delta ^{(2)}) \\ 
&=& \rho^2 \mathrm{Cov}(\mathbf{y}^{(1)},\mathbf{y}^{(1)}) + \mathrm{Cov}(\delta ^{(2)},\delta ^{(2)}) = \rho^2 K_{1} + K_{2}
\eea 
that is, $\mathbf{y}^{(2)} \sim \mathcal{N}(0, \rho^2 K_{1} + K_{2})$.

The most important term in multi-fidelity theory is the covariance between $\mathbf{y}^{(1)}$ and $\mathbf{y}^{(2)}$, which represents the inherent correlation between data sets with different levels of fidelity and is derived as $\mathrm{Cov}(\mathbf{y}^{(1)}, \mathbf{y}^{(2)}) = K_{12} = \rho \mathrm{Cov}(\mathbf{X} ,\mathbf{X} ) = \rho K_{1}$ due to the same independence restriction. Now the multi-fidelity structure can be written in the following compact form of a multivariate Gaussian process:
\begin{equation} \label{eq:joint_distr}
\begin{pmatrix} \mathbf{y}^{(1)} \\
\mathbf{y}^{(2)} \end{pmatrix} \sim \mathcal{N}\Biggl(0,\begin{pmatrix} K_{11} & K_{12}\\
K_{21} & K_{22}
 \end{pmatrix}\Biggr),
\end{equation}
where $K_{11}=K_1$, $K_{22} \neq K_2, K_{12} = K_{21}$ due to symmetry.
The importance of $\rho$ is quite evident from the term $K_{12}$; specifically, when $\rho = 0$, the high fidelity and low fidelity models are completely decoupled and there will be no improvements of the prediction at all by combining the two models.

The next step is to make prediction of $\mathbf{y}_*^{(2)}$ given the corresponding input vector $\mathbf{x}_* $, two levels of training data \{$\mathbf{X} , \mathbf{y}^{(1)}$\} and \{$\mathbf{X} ,\mathbf{y}^{(2)}$\}. To this end, we first write down the following joint density:
\begin{equation} \label{eq:joint_distr2}
\begin{pmatrix} \mathbf{y}_*^{(2)} \\
 \mathbf{y}^{(1)} \\
\mathbf{y}^{(2)} \end{pmatrix} \sim \mathcal{N}\Biggl(0,\begin{pmatrix} K_{**} & K_{*1} & K_{*2}\\
K_{1*} & K_{11} & K_{22} \\
K_{2*} & K_{21} & K_{22}
 \end{pmatrix}\Biggr),
\end{equation}
where $K_{**} =  \rho^2 K_{1}^* + K_{2}^*$, $K_{*1} = \rho K_{1}^*$ with $K_1^* = K_1(\mathbf{X}_* ,\mathbf{X}_* ) = 
\mathrm{Cov}(\mathbf{y}^{(1)}_*, \mathbf{y}_*^{(1)})$ and $K_2^* = K_2(\mathbf{X}_* ,\mathbf{X}_* ) = 
\mathrm{Cov}(\delta^{(2)}_*, \delta_*^{(2)})$; then following similar procedures as in section~\ref{sec:GPR}, the final predictive distribution of $\mathbf{y}_*^{(2)}|\mathbf{X}_* , \mathbf{X}, \mathbf{y}^{(1)}, \mathbf{y}^{(2)}$ is again a Gaussian $\mathcal{N}(\bar{\mathbf{y}}_*^{(2)}, \mathrm{Var})$, where
\bea 
\bar{\mathbf{y}}_*^{(2)} = K_* K^{-1}Y,~\mathrm{Var} = K_{*}K_{*}^{\top} - K_{*}K^{-1}K_{*}^{\top},\\
Y = \begin{pmatrix} \label{eq:fidelity_kernel} \mathbf{y}^{(1)}\\ \mathbf{y}^{(2)}\end{pmatrix},~
K_* = \begin{pmatrix} K_{*1} & K_{*2} \end{pmatrix},~
K = \begin{pmatrix} K_{11} & K_{12} \\ K_{21} & K_{22} \end{pmatrix}.
\eea

We note in passing that since there are two correlations function $K_1$ and $K_2$, two sets of hyper-parameters regarding the kernel width and an extra scaling parameter $\rho$ have to be optimized following the similar approach as explained in section (\ref{sec:LC}). This algorithm has already successfully been applied to the prediction of band gaps of elpasolite compounds with high accuracy~\citep{pilania2017}. But it can be naturally extended to other properties. So far, not much work has been done using this algorithm, its potential to tackle complicated chemical problems has yet to be unraveled by future work.

\subsection{$\Delta$-Machine learning}
A naive version of multi-fidelity learning is the so called $\mathrm{\Delta}$-machine learning model. Its performance is useful for the prediction of various molecular properties~\citep{DeltaPaper2015}. In this model, $N_1$ is equal to $N_2$, the low and high-fidelity models are respectively called baseline and target. The baseline property ($y^{(b)}$) is associated with baseline geometry as encoded in its representation $\vec{x}^{(b)}$), and target property $y^{(t)}$ is associated with target eometry $\vec{x}^{(t)}$, respectively. The workhorse of this model is 
\beq
y^{(t)}_* = y^{(b)}_* + \sum_{i=1}^N c_i k(\vec{x}^{(b)}_*, \vec{x}^{(b)}_i)
\eeq 
Note that we did not use the target geometry at all for the reason that 1) it is expensive to calculate; 2) it is not necessary for the test molecules. 

The $\Delta$-ML model has been shown to be capable of yielding highly accurate results for energies if a proper baseline model is used. Other properties can also be predicted with much higher precision compared to traditional single fidelity model~\citep{DeltaPaper2015}. What is more, this approach can save substantial computational time. However, the $\Delta$-machine learning model is not fully consistent with the multi-fidelity model. The closest scenario is that we set $K_1 = K_2$ when evaluating kernel functions in equation (\ref{eq:fidelity_kernel}), but this will result in something still quite different. There are further issues one would like to resolve, including that (i) the coupling between different fidelities is not clear and that the correlation is rather naively accounted for through the $\Delta$ of the properties from two levels, assuming a smooth transition from one property surface (e.g., potential energy surface) from one level of theory to another. This is questionable and may fail terribly in some cases; (ii) it requires the same amount of data for both levels, which can be circumvented by building recursive versions.

\section{Representation} \label{sec:Repr}

The problem of how to represent a molecule or material has been a topic dating back to many decades ago and the wealth of information (and opinions) about this subject is well manifested by the collection of descriptors compiled in Todeschini and Consonni's Handbook of molecular descriptors~\citep{todeschini2008}. According to these authors, the molecular descriptor is defined as ``the final result of a logic and mathematical procedure which transforms chemical information encoded within a symbolic representation of a molecule into a useful number or the result of some standardized experiment". Whilst the majority of these descriptors are graph-based and used for quantitative structure and activity relationships (QSAR) applications (typically producing rather rough correlation between properties and descriptor), our focus is on QML models, i.e.~physics based, systematic and universal predictions of well-defined quantum mechanical observables, such as the energy~\cite{QMLessayAnatole}. Thus, to better distinguish the methods reviewed here-within from QSAR, we prefer to use the term ``representation'' rather than ``molecular descriptor''. Quantum mechanics offers a very specific recipe in this regard: A chemical system is defined by its Hamiltonian which is obtained from elemental composition, geometry, and electron number exclusively. As such, it is straightforward to define the necessary ingredients for a representation: It should be some vector (or fingerprint) which encodes the compositional and structural information of a given neutral compound.

\subsection{The essentials of a good representation} \label{sec:GoodRepr}
There are countless ways to encode a compound into a vector, but what representation can be regarded as ``good''? Practically, a good representation should lead to a decent learning curve, i.e., error steadily decreases as a function of training set size. Conceptually, it should fulfill several criteria, including primarily uniqueness (non-ambiguity), compactness and being size-extensive~\citep{OAvL_FRD}.

Uniqueness (or being non-ambiguous) is indispensable for ML models. We consider a representation to be unique if there is no pair of molecules that produces the same representation. Lack of uniqueness would results in serious consequences, such as ceasing to learn at an early stage or no learning at all from the very beginning. The underlying origin is not hard to comprehend. Consider two representation vectors $\mathbf{x}_1$ and $\mathbf{x}_2$ for two compounds associated with their respective properties $y_1$ and $y_2$. Now suppose $\mathbf{x}_1 = \mathbf{x}_2$ while $y_1 \neq y_2$ (no degeneracy is assumed). One extreme case is that only these two points are used when training the ML model, obviously we will encounter a singular kernel matrix with all elements being 1; huge prediction errors will result and basically there is no learning. Even if molecules like these are not chosen for training it should be clear that such a representation introduces a severe and systematic bias. Furthermore, when trying to predict $y_1$ and $y_2$ after training, the estimate will be the same as the input to the machine is the same. The resulting test error is therefore directly proportional to their property difference.

The compactness requires atom index permutation, rotational and translational invariance, i.e., all redundant degrees of freedom of the system should be removed as much as possible while retaining the uniqueness. This can lead to a more robust representation, meaning 1) the size of training set needed may be significantly reduced; 2) the dimension of the representation vector (thus the size) is minimized, a virtue which becomes important when the necessary training set size becomes large. 


Being size-extensive is crucial for prediction of extensive properties, among which the most important, the energy. This leads to the so-called atomic representation or local representation of an atom in a compound. The local unit atom can also consist of bonds, functional groups or even larger fragments of the compound. As pointed out in section \ref{sec:GPR_local}, this type of representation is the crucial stepping stone for building scalable machine learning models. Even intensive properties such as HOMO-LUMO gap which typically do not scale with system size, can be modeled within the framework of atomic representations, as illustrated using the Re-Match metric~\citep{Sandip2016}. 
For specific problems, such as force predictions, an analytic form  of representation is desirable for analysis and rapid evaluation, and for subsequent differentiation (with respect to nuclear charges and coordinates) so as to account for response properties.


\subsection{Rational design} 
It is not obvious how to obtain an optimal representation. In order to obtain a good representation, one has to gain intensive knowledge about the system and structure-property relationship. Use of simplified approximations to solutions of Schr\"odinger's equation are particularly powerful. The most approximative, yet atomistic, models of SE are universal force fields (FF) which typically reproduce the essential physics for certain system classes, such as bio-organic molecules, reasonably well. Namely, the atom-pairwise two-body interactions in force-fields typically decay as $1/R^n$ ($R$ being the internuclear distance and $n$ being some integer), while 3- and 4-body parts behave as periodic functions of angle and dihedral angle (modern force field approaches also include 2- to ($n-1$)-body interaction in $n$-body interactions). FFs are essentially a special case of the more general many-body expansion (MBE) in interatomic contributions, i.e., an extensive property of the system (e.g., total energy) is expanded in a series of many-body terms, namely, 1-, 2- and 3-body terms, $\cdots$, i.e.,
\begin{equation} \label{eq:MBE}
\begin{split}
E(\{ \mathbf{R}_I \}) &= \sum_{I}^{[Z]} E^{(1)}(R_I) + \sum_{J > I}^{[Z]} E^{(2)}(R_{IJ}) + \sum_{\substack{K>J>I}}^{[Z]} E^{(3)}(R_{IJ},R_{IK},\theta_{IJK}) + \cdots 
\end{split}
\end{equation}
where $E^{(n)}$ is the $n$-body interaction energy, $R_{IJ}$ is the interatomic distance between atom $I$ and $J$, $\theta_{IJK}$ is the angle spanned by two vectors $\vec{R}_{IJ}$ and $\vec{R}_{IK}$. Other important properties can also be expressed in a similar fashion.

By utilizing the basic variables in MBE, including distance, angles and dihedral angles in their correct physics based functional form (for instance, the aforementioned $1/R^n$ dependence of 2-body interaction strength) one can already build some highly efficient representations such as BAML and SLATM ({\em vide infra}). This recipe relies heavily on pre-conceived knowledge about the physical nature of the problem. 

\subsection{Numerical optimization} 

It is possible that for some systems and properties, one does not know which features are of primary importance. And it is not an option to try all features one-by-one considering that there are so many possibilities. In such a situation, the least absolute shrinkage and selection operator (LASSO) can offer suitable relief. LASSO  is basically a regression analysis method. Consider a simple linear model: the property of a system is a linear functions of its features, i.e., $\mathbf{y=Xc}$, where $\mathbf{X}$ is a matrix with each of the $N$ rows being the descriptor vector $\mathbf{x}_i$ of length $D$ for each training data points, $\mathbf{c}$ is the $D$-dimensional vector of
coefficients, and $\mathbf{y}$ is the vector of training properties with the $i$-th property being $y_i$. Our task is to find the tuple of features that yields the smallest sum of squared error: $||\mathbf{y-Xc}||_2^2$. Within LASSO, it is equivalent to a convex optimization problem, i.e., 
\beq
\underset{ \mathbf{c} \in {\rm I\!R}^D}{ \mathrm{argmin} }~||\mathbf{y-Xc}||_2^2 + \lambda ||\mathbf{c}||_1
\eeq
where the use of $L_1$ norm of regularization term is pivotal, i.e., smaller $L_1$ norm can be obtained when larger $\lambda$ is used, thereby purging features of lesser importance. This approach has been exemplified for the prediction of relative crystal phase stabilities (rock-salt vs. zinc-blende) in a series of binary solids~\citep{descriptor_design_LASSO}. Unfortunately, this approach is limited in that it works best for rather low-dimensional problems. Already for typical organic molecules, the problem becomes rapidly intractable due to coupling of different degrees of freedom. Under such circumstances, it appears to be more effective to adhere to the aforementioned rational design based heuristics, as manifested by the fact that almost all of the \textit{ad-hoc} representations in the literature are based on manual encoding. 

\subsection{An overview of selected representations}
Over the years, numerous molecular representations have been developed by several research groups working on QML. It's not our focus to enumerate all of them, but to list and categorize the popular ones. Two categories are proposed, one is based on many-body expansions in vectorial or tensorial form, such as Coulomb matrix (CM), Bag of Bonds (BoB), Bond, Angle based Machine Learning (BAML), Spectrum of London and Axilrod-Teller-Muto potential (SLATM), and the alchemical and structural radial distribution based representation introduced by Faber, Christensen, Huang, and von Lilienfeld (FCHL). The other category is an electron density model based representation called Smooth Overlap of Atomic Positions (SOAP). 

\subsubsection{Many-body potential based representation}
The Coulomb matrix (CM) representation was first proposed in the seminal paper by~\citep{CM}. It is a square atom-by-atom matrix with off diagonal elements corresponding to the nuclear Coulomb repulsion between atoms, i.e., $\mathrm{CM}_{IJ} = Z_I Z_J/R_{IJ}$ for atom index $I \neq J$. Diagonal elements approximate the electronic potential energy of the free atom, and is encoded as $-0.5Z_I^{2.4}$. To enforce invariance of atom indexing, one can sort the atom numbering such that the sum of $L_2$ and $L_1$ norm of each row of the Coulomb matrix descends monotonically in magnitude. Symmetrical atoms will result in the same magnitude. A slight improvement over the original CM can be achieved by varying the power low of $R_{IJ}$~\cite{baml}. Best performance is found for an exponent of 6, reminiscent of the leading order term in the dissociative tail of London dispersion interactions. Thus, the resulting representation is also known as London matrix (LM). The superiority of LM is attributed to a more realistic trade-off between the description of more localized covalent bonding and long-range intramolecular non-covalent interactions~\citep{baml}.


\begin{figure*} \label{fig:homometric}
  \centering
  \includegraphics[width=0.5\columnwidth]{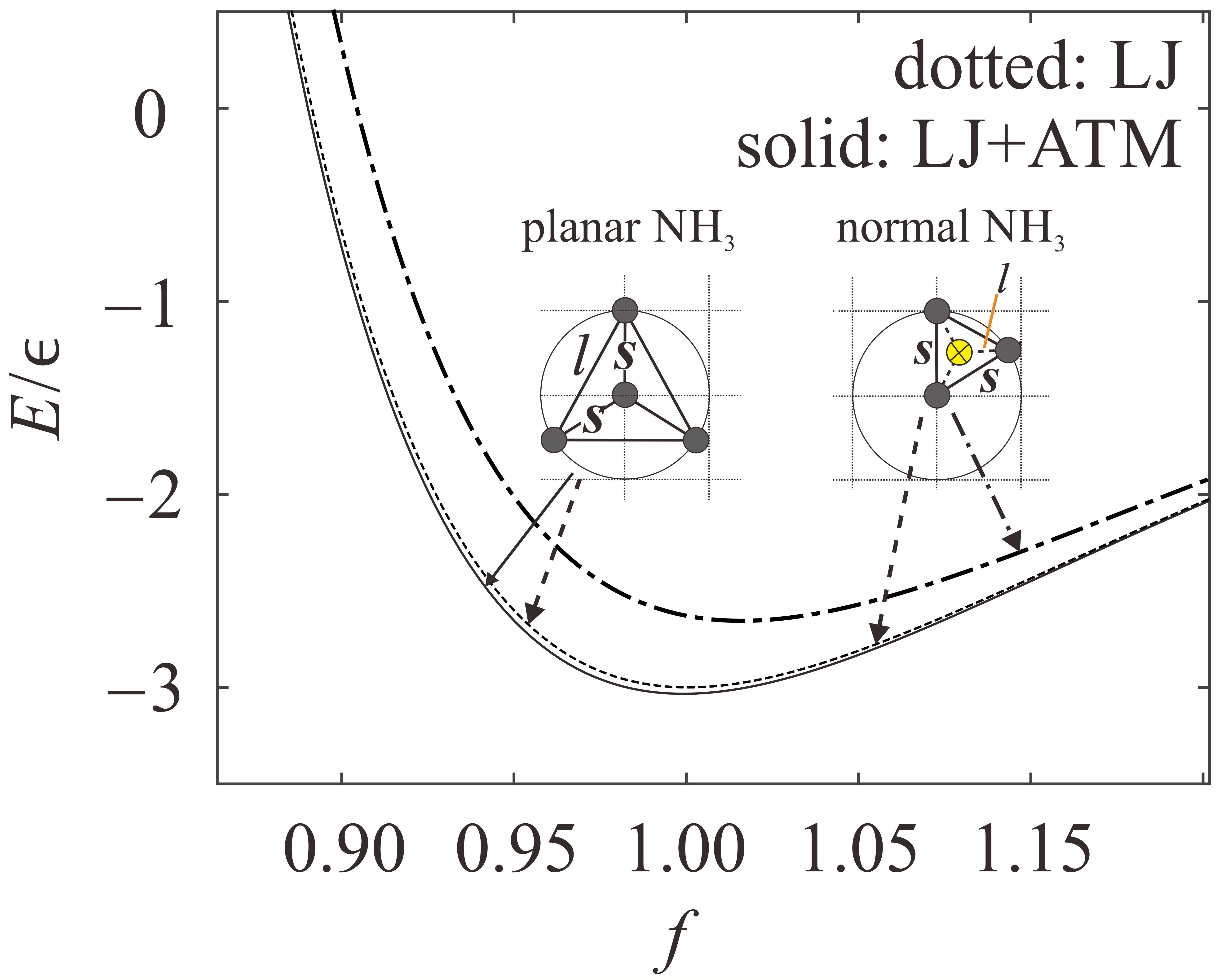}
    \caption{Two body interaction is not enough to capture the physics of a pair of homometric molecules. In the figure, the energy of the two molecules are approximated as summation of LJ potentials with (dashed lines) or without 3-body ATM potentials (solid line) and plotted as a function of $f$, the scaling factor of all coordinates of the two molecules. LJ, ATM stands for Lennard-Jones and Axilrod-Teller-Muto vdW potential, respectively. The letters $s$ and $l$ labels the two existing different bond lengths, standing for `short' and `long'. The atom represented by a yellow filled circile with cross means out of plane.}
  \end{figure*}

In spite of the great virtue of uniqueness encoded in CM, it generally suffers from a high offset of learning curve~(see Fig. 3).
In contrast, the bag-of-bond (BoB) representation~\citep{BoB}, a bagged (vectorial) stripped down version of the CM, turns out to result in learning curves with lower off-set than CM (see Fig. 3). 
The BoB representation is a 1-D array, constructed as the concatenation of a series of bags (1-D arrays as well), each corresponds to a specific type of atomic pair, e.g., all C-O pairs (covalently and non-covalently bonded) in the molecule are grouped into the bag labeled as CO; similarly for all other combinations of elemental pairs. Each bag thus includes a set of nuclear Coulomb repulsion values. Each bag is then sorted in descending order. In cases that the same type of bag for two molecules has not the same size the smaller bag is padded with zeros. Through bagging the performance is improved in comparison to the CM matrix. But inevitably, crucial higher-order information, such as the angular part, is missing. Due to its exclusive reliance on sorted two-body terms, BoB is not a unique representation, as also manifested by the deterioration of its slope in the learning curve for large training set sizes (see Fig. 3). 
This loss of information can also be illustrated for a pair of homometric molecules (same atom types, same set of interatomic distances) as displayed in Fig~2. If we make a plot of the potential energy (approximated as a sum of  Lennard-Jones potentials) curve of both planar and tetrahedral molecules as a function of the scaling factor $f$ of all coordinates, we will end up with the same curve due to a spurious degeneracy imposed by lack of uniqueness. The BoB representation would not distinguish between these two molecules. Only after addition of higher order many-body potential terms (e.g., the 3-body Axilrod-Teller-Muto  potential), the spurious degeneracy is lifted. 



\begin{figure}\label{fig:performance}
  \centering
  \includegraphics[width=0.8\columnwidth]{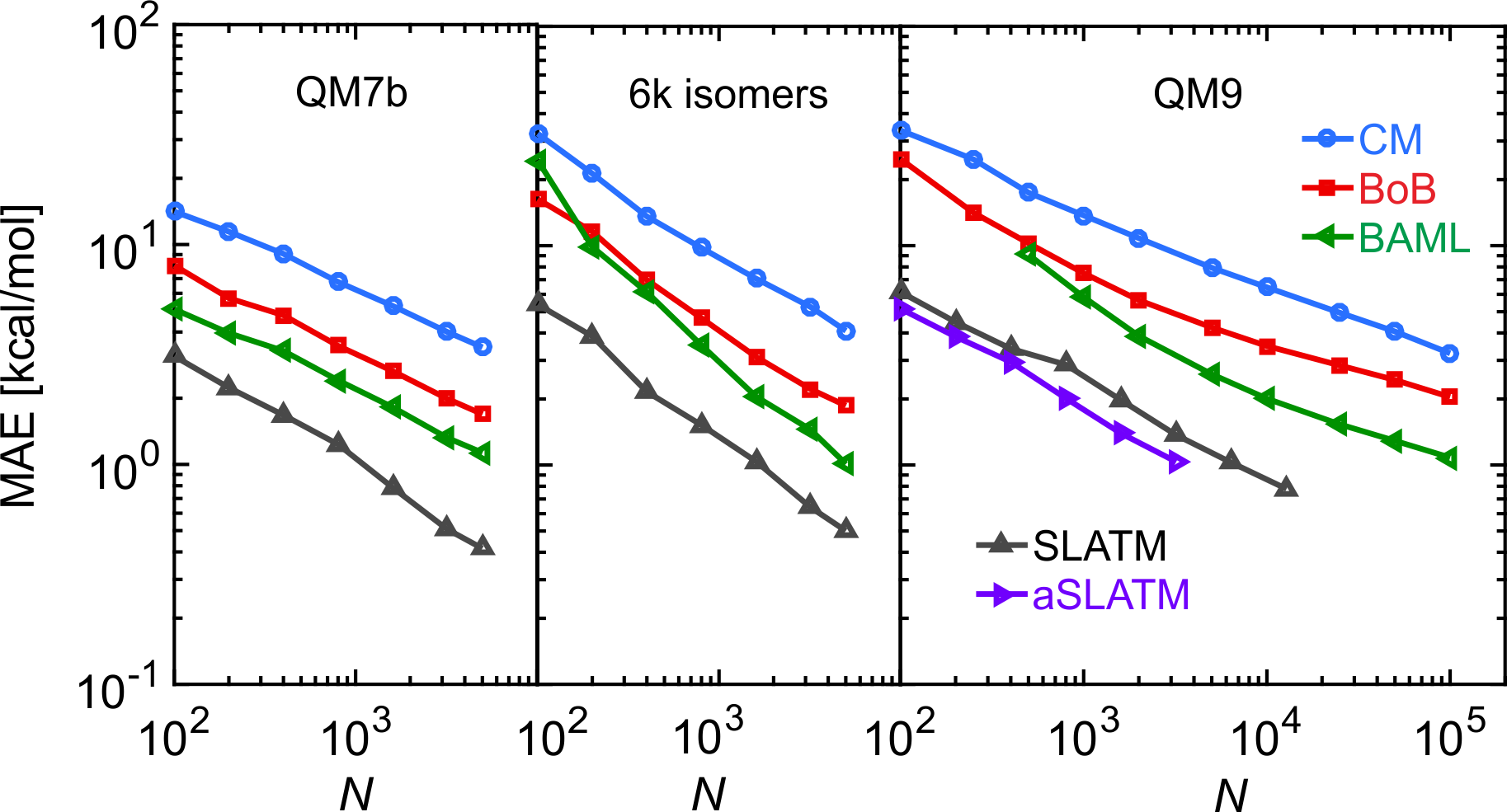}
  \caption{Comparison of the learning curves for different representations for three typical datasets (QM7b~\citep{CM}, QM9~\citep{gdb9_data,gdb17} and 6k isomers from QM9). Note that the size and composition for molecules in all the three datasets are comparable, i.e., the dimensionality $d$'s of these systems are similar, hence almost the same learning rates is observed for all representations with no (or less) suffer from uniqueness issue. For QM7b dataset, a much lower offset is shown as the relevant molecules are much more relaxed than those in QM9 and 6k isomers, thus given any representation, its target similarity is larger for this dataset compared to others.}
\end{figure}

Based on this simple example, an important lesson learned is that collective effects which go beyond pairwise potentials are of vital importance for the accurate modeling of fundamental properties such as energies. While adhering to the ideas of bagging for efficiency, a representation consisting of extended bags can be constructed, each may contain interatomic interaction potentials up to 3- and 4-body terms. BAML was formulated in this way, where 1) all pairwise nuclear repulsions are replaced by Morse/Lennard-Jones potentials for bonded/non-bonded atoms respectively; 2) The inclusion of 3- and 4-body interactions of covalently bonded atoms is achieved using periodic angular and torsional terms, with their functional form and parameters extracted from the Universal Force Field (UFF)~\citep{baml,uff}. BAML achieves a noticeable boost of performance when compared to BoB or CM. Interestingly, the performance is systematically improving upon inclusion of higher and higher order many-body terms, as the proposed energy model is getting more and more realistic, i.e., increasing similarity to target. Meanwhile, and not surprisingly the uniqueness issue, existing in two-body representations such as BoB, is also resolved (see Fig.~\ref{fig:performance}).
The main drawback of BAML, however, is that it requires pre-existing force fields, implying a severe bias when it comes to new elements or bonding scenarios. It would therefore be desirable to identify a representation which is more compact and \textit{ab-initio} in nature. 


The so-called SLATM representation~\citep{huang2018} enjoys all these attributes. It has two variants: a local and a global one. The basic idea of SLATM is to represent an atom indexed $I$ in a molecule by accounting for all possible interactions between atom $I$ and its neighboring atoms through many-body potential terms multiplied by a normalized Gaussian distribution centered on the relevant variable (distance or angle). So far, 1-, 2- and 3-body terms have been considered. The 1-body term is simply represented by the nuclear charge, while the two-body part is expressed as 
\beq \label{eq:bop}
\frac{1}{2} Z_I \sum_{J\neq I} Z_J\delta(\mathbf{r}-\mathbf{R}_{IJ})g(\mathbf{r})
\eeq 
where $\delta(\cdot)$ is set to normalized Gaussian function $\delta(x) = \frac{1}{ \sigma \sqrt{2\pi} }e^{ - x^2 }$,
$g(r)$ is a distance dependent scaling function, capturing the locality of chemical bond and chosen to correspond to the leading order term in the dissociative tail of the London potential $g(R) = \frac{1}{R^6}$. The 3-body distribution reads
\beq \label{eq:bot}
\frac{1}{3} Z_I \sum_{J\neq K \neq I} Z_J Z_K \delta(\theta - \theta_{IJK})h(\theta, \mathbf{R}_{IJ}, \mathbf{R}_{IK}) 
\eeq
where $\theta$ is the angle spanned by vector $\mathbf{R}_{IJ}$ and $\mathbf{R}_{IK}$ (i.e.,$\theta_{IJK}$) and treated as a variable. $h(\theta, \mathbf{R}_{IJ}, \mathbf{R}_{IK})$ is the 3-body contribution depending on both internuclear distance and angle, and is chosen in form to model the Axilrod-Teller-Muto~\cite{atm,atm2} vdW potential
\bea 
h(\theta, \mathbf{R}_{IJ}, \mathbf{R}_{IK}) &=& \frac{1+\cos \theta \cos \theta_{JKI}\cos \theta_{KIJ}}{(R_{IJ}R_{IK}R_{KJ})^3} 
\eea
Now we can build the atomic version aSLATM for an atom $I$ through concatenation of all the different many-body potential spectra involving atom $I$ as displayed in equation (\ref{eq:bop}) and (\ref{eq:bot}). As for the global version SLATM, it simply corresponds to the sum of the atomic spectra. 

SLATM and aSLATM outperforms all other representations discussed so far, 
as evinced by  learning curves  shown in Fig.~\ref{fig:performance}. 
This outstanding performance is due to several aspects: 1) almost all the essential physics in the systems is covered, including the locality of chemical bonds as well as many-body dispersion; 2) the inclusion of 3-body terms significantly improves the learning. 3) the spectral distribution of radial and angular feature now circumvents the problem of sorting within each feature bag, allowing for a more precise match of atomic environments.

Most recently, the FCHL representation has been introduced~\citep{faber2017alchemical}. It amounts to a radial distribution in elemental and structural degrees of freedom. The configurational degrees of freedom are expanded up to three-body interactions. Four-body interactions were tested but did not result in any additional improvements. For known data-sets FCHL based QML models reach unprecedented predictive power and even outperform aSLATM and SOAP (see below). In the case of the QM9 dataset, for example, FCHL based models of atomization energies reach chemical accuracy after training on merely $\sim$1'000 molecules. 

\subsubsection{Density expansion based representation} 

Within the Smooth Overlap of Atomic Positions (SOAP)~\citep{soap_2013} idea of a representation, an atom $I$ in a molecule is represented as the local density of atoms around $I$. Specifically, it is represented by a sum of Gaussian functions with variance $\sigma^2$ within the environment (including the central atom $I$ and its neighboring atoms $Q$'s), with the Gaussian functions centered on $Q$'s and $I$:
\beq \label{eq:adens}
\rho_{I}(\mathbf{r}) = \sum_{Q} \exp \left ( -\frac{(\mathbf{r} - \mathbf{R}_Q)^2}{2\sigma^2} \right )
\eeq
where $\mathbf{r}$ is the vector from the central atom $I$ to any point in space, while $\mathbf{R}_Q$ is the vector from atom $I$ to its neighbour $Q$.
The overlap of $\rho_I$ and $\rho_J$ then can be used to calculate a similarity between atoms $I$ and $J$. However, this similarity not rotationally invariant. To overcome this, we can integrate out the rotational degrees of freedom for all 3-dimensional rotations $\hat{R}$, and thus the SOAP kernel is defined, 
\beq \label{eq:soap_kernel}
\tilde{k}(I,J) = \int d \hat{R} \left | \int d\mathbf{r} \rho_I(\mathbf{r}) \rho_J(\hat{R} \mathbf{r}) \right |^2,
\eeq 
To enforce the self-similarity to be normalized, the final SOAP similarity measure takes the form of
\beq \label{eq:soap2}
k(I,J) = \frac{\tilde{k}(I,J)}{\sqrt{\tilde{k}(I,I)\tilde{k}(J,J)}}
\eeq

The integration in equation (\ref{eq:soap_kernel}) can be carried out by first expanding $\rho_I (\mathbf{r})$ in equation (\ref{eq:adens}) in terms of a set of basis functions composed of orthogonal radial functions and spherical harmonics and then collect the elements in the rotationally invariant power spectrum, based on which $k$ can be easily calculated. The interested reader is referred to \citep{soap_2013}.

SOAP has been used extensively and successfully to model systems such as silicon bulk or water clusters, each separately with many configurations. These elemental or binary systems are relatively simple as the diversity of chemistries encoded by the atomic environments is rather limited. A direct application of SOAP to molecules where there are substantially more possible atomic environments, however, yields learning curves with rather large off-sets. This is not such a surprise, as essentially the capability of atomic densities to differentiate between different atom pairs, atom triples, and so on, is not so great. This shortcoming remains even if one treats different atom pairs as different variables, as was adopted in \citep{Sandip2016}; averaging out all rotational degrees of freedom might also impede the learning progress due to loss of relevant information.
To amend some of these problems, a special kernel, the RE-Match kernel \citep{Sandip2016}, was introduced. And most recently, combining SOAP with a multi-kernel expansion enabled additional improvements in predictive power~\citep{CeriottiScienceUnified2017}. 

\section{Training set selection}
The last section of this chapter deals with the question of how to select training sets. The selection procedure can have a severe effect on the performance. The predictive accuracy appears to be very sensitive on how we sample the training molecules for any given representation (or better ones). Training set selection can actually be divided into two parts: (1) how to create training set. The general principle is that the training set should be representative, i.e.~it follows the same distribution as all possible test molecules in terms of input and output. This will formally prevent extrapolation and thereby minimize prediction errors. 
(2) how to optimize the training set composition. 

The majority of algorithms in literature deal with (2), assuming the existence of some large dataset (or a dataset trivial to generate) from which one can draw using algorithms such as ensemble learning, genetic evolution, or other  ``active learning'' based procedures\citep{Shapeev2017}. All of these methods have in common that they select the training set from a given set of configurations based only on the unlabeled data. 
This is particularly useful for ``learning on the fly'' based {\em ab initio} molecular dynamics simulations~\cite{lotf2004}, where expensive quantum-mechanical calculation are carried out only when the configurations are sufficiently ``new''.

Step 1 stands out as a challenging task and few algorithms are competent. The most ideal approach is of course an algorithm that can do both parts within one step, the only competent method we know is the ``amons'' approach. We will elaborate on all these concepts below.

\subsection{Genetic optimization}

To the best of our knowledge, the first application of a GA for generation and study of optimal training set compositions for QML model was published in \citep{geneticAlgorithm_trainingSet}. The central idea of this approach is outlined as follows. For a given set ($S_0$) containing overall $N$ molecules, the GA procedure consists of three consecutive steps to obtain the ``near-optimal'' subset of molecules from $S_0$ for training the ML model~\citep{geneticAlgorithm_trainingSet}: (a) randomly choose $N_1$ molecules as a trial training set $s_1$; repeat $M$ times. This forms a population of training sets, termed the parent population and labeled as $\hat{s}^{(1)} = \{s_1, s_2, \dots, s_M\}$. (b) An ML model is trained on each $s_i$, and then tested on a fixed set of out-of-sample molecules, resulting in a mean prediction error $e_i$, which is assigned to $s_i$ as a measure of how fit $s_i$ is as the ``near-optimal'' training set and dubbed ``fitness'' . Therefore, the smaller $e_i$ is, the larger the fitness is. (c) $\hat{s}^{(1)}$ is consecutively evolved through selection (to determines which $s_i$'s in $\hat{s}^{(1)}$ should remain in the population to produce a temporarily refined smaller set $\hat{t}^{(1)}$; a set $s_i$ with larger fitness means higher probability to be kept in $\hat{t}^{(1)}$), crossover (to update $\hat{s}^{(1)}$ from $\hat{t}^{(1)}$ and the new $\hat{s}$ is labeled as $\hat{s}^{(2)}$ with each set $s_i$ in $\hat{s}^{(2)}$ obtained through mixing the molecules from two $s_i$'s in $\hat{t}^{(1)}$), mutation (to change molecules in some $s_i$'s in $ \hat{s}^{(2)}$ randomly to promote diversity in $\hat{s}^{(2)}$, e.g., replace -CH$_2$- fragment by -NH- for some molecule. (d) Go to step (b) and repeat the process until there is no more change in the population and the fitness ceases to improve .
We label the final updated trial training set as $\hat{s}$.

It's obvious that the molecules in $\hat{s}$ should be able to represent all the typical chemistry in all molecules in $S_0$, such as linear, ring, cage-like structure and typical hybridization states ($sp,sp^2,sp^3$) if they are abundant in $S_0$. Once trained on $\hat{s}$, the ML model is guaranteed to yield typically significantly better results as the fitness is constantly increasing. This is not useful since the GA ``tried'' this already; the usefulness has to be assessed by the generalizability of $\hat{s}$ as training set to test on a new set of molecules not seen in $S_0$. Indeed, as shown in \citep{geneticAlgorithm_trainingSet}, 
significant improvements in off-sets can be obtained when compared to random sampling. 
While the remaining out-of-sample error is still substantial, this is not surprising due to the use of less advantageous representations. 
One of the key-findings in this study were that upon genetic optimization (i) the distance distributions between training molecules were shifted outward, and (ii) the property distributions of training molecules were fattened.

\subsection{Amons} 
We note that the naive application of active learning algorithms will still result in QML models which suffer from lack of transferability, in particular when it comes to the prediction of larger compounds or molecules containing chemistries not present in the training set. Due to the size of chemical compound space this issue still imposes a severe limitation for the general applicability of QML. These problems can, at least partially, be overcome by exploring and exploiting the locality of an atom in molecule~\citep{huang2018}, resulting from the nearsightedness principle in electronic systems~\citep{nearsightedness,stjin2017}. 

We consider a valence saturated query molecule for illustration, for which we try to build an ``ideal'' training set. As is well known, any atom $I$ (let us assume a $sp^3$ hybridized C) in the molecule is characterized by itself and its local chemical environment. To a first order approximation, we may consider its coordination number (\textit{CN} for short) to be a distinguishing measure of its atomic environment and we can roughly say that any other carbon atom with a coordination number of 4 is similar to atom $I$, as their valence hybridization states are all $sp^3$. Another carbon atom with $CN$ = 3 in hybridization state of $sp^2$ would be significantly different compared to atom $I$. It is clear, however, that \textit{CN} as an identifier of atomic environment type is not enough: An $sp^3$ hybridized C atom in methane molecule (hereafter we term it as a genuine C-$sp^3$ environment) is almost purely covalently bonded to its neighbors, while in CH$_3$OH, noticeable contributions from ionic configurations appear in the valence bond wavefunction due to the significant electronegativity difference between C and O atoms. Thus one would expect very different atomic properties for the $sp^3$-C atoms in these two environments as manifested, for instance, in their atomic energy, charge, or $^{13}$C-NMR shift. Alternatively, we can say that oxygen as a neighboring atom to $I$ has perturbed the ideal $sp^3$ hybridized C to a much larger extent in CH$_3$OH than the H atom has in methane. To account for these differences, we can simply include fragments which contain $I$ as well as all its neighbors. Thus we can obtain a set of fragments, for each of which the bond path between $I$ and any other atom is 1. 

Extending this kind of reasoning to the second neighbor shell, we can add new atoms with a bond path of 2 relative to atom $I$ in order to account for further, albeit weaker, perturbation to atom $I$. As such, we can gradually increase the size of included fragments (characterized by the number of heavy atoms) until we believe that all effects on atom $I$ have been accommodated. The set of unique fragments can then be used as a training set for a fragment based QML model. Note that we saturate all fragments by hydrogen atoms. 
These fragments can be regarded as effective quasi-atoms which are defined as \underline{a}tom in \underline{m}olecule, or ``am-on''. Since amons repeat throughout chemical space, they can be seen as the ``words'' of chemistry (target molecules being ``sentences'') or as ``DNA'' of chemistry (target molecules being genes and properties their function). 
Given the complete set of amons, any specific, substantially larger, query molecule can be queried. Used in conjunction with an atomic representation such as aSLATM or FCHL, amons enable a kind of chemical extrapolation which holds great promise to more faithfully and more efficiently explore vast domains chemical space~\citep{huang2018}.



\begin{figure*} \label{fig:amon_vs_random}
  \centering
  \includegraphics[width=0.6\columnwidth]{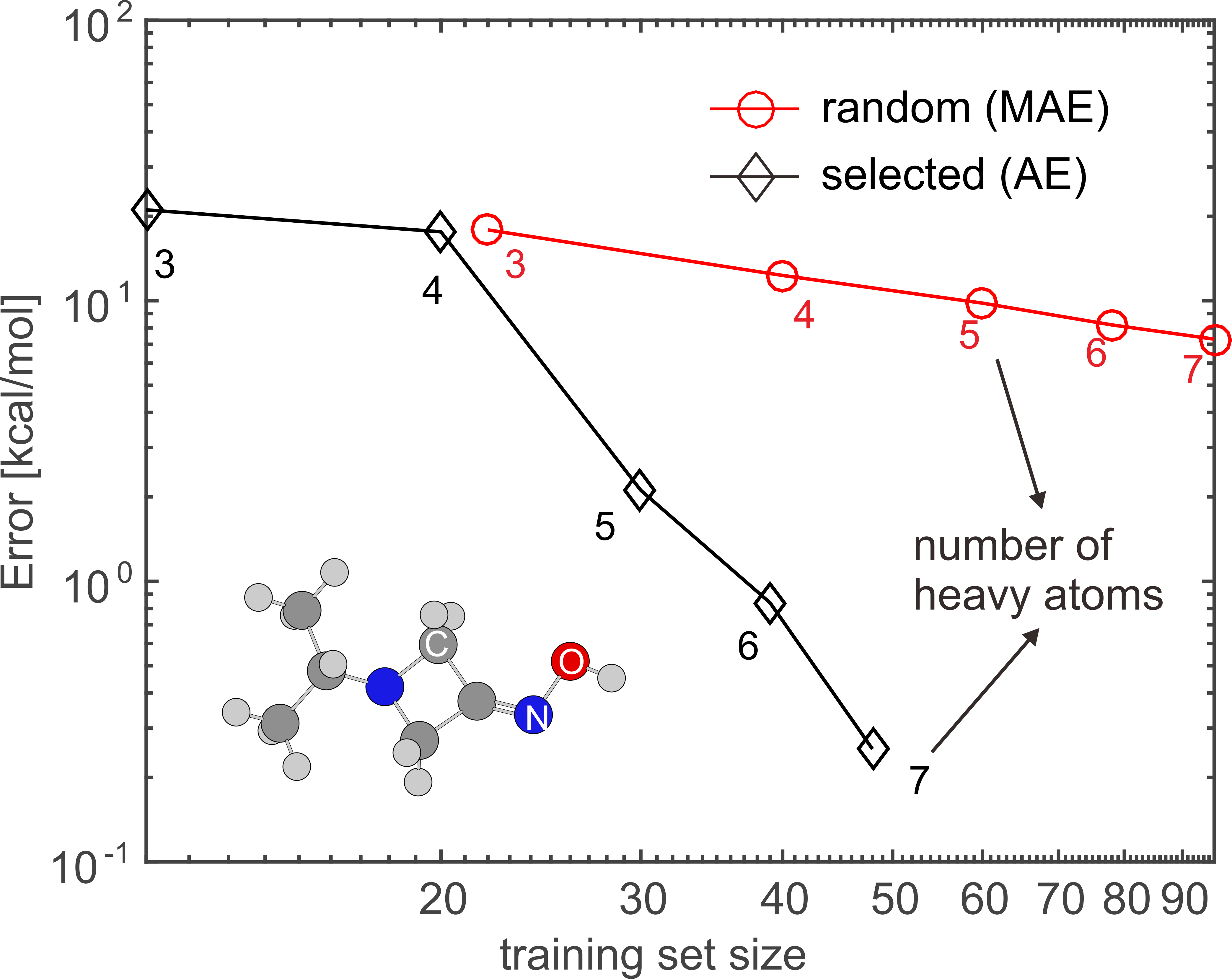}
  \caption{Comparison of the learning curves obtained for one molecule (see the inset) from random selection of training set and amons, respectively. For each red scatter point, errors were averaged over 100 random samplings.}
\end{figure*}

To demonstrate the power of amons, we show the example of predicting the potential energy of a molecule present as an inset in Fig.~\ref{fig:amon_vs_random}. With amons as the training set, chemical accuracy (1 kcal/mol) is reached after training on only 40 amons (with amons being not larger than 6 heavy atoms). Sampling amons at random, the slope of the learning curve is substantially worse. 

\section{Conclusion}

We have discussed primarily the basic mathematical formulations of all typical ingredients of quantum machine learning (QML) models which can be used in the context of quantum mechanical training and testing data. We explained and reviewed why ML models can be fast and accurate when predicting quantum mechanical observables for out-of-sample compounds. It is the authors' opinion that QML can be seen as a very promising approach, enabling the exploration of systems and problems which hitherto were not amenable to traditional computational chemistry methods. 

In spite of the significant progress made within the last few years, the field QML is still very much in a stage of infancy. This should be clear when considering that the properties that have been explored so far are rather limited and relatively fundamental. The primary focus has been on ground state or local minimum properties.  Application to excited states still remain a challenge~\citep{ML-TDDFTEnrico2015}, just as well as conductivity, magnetic properties, or phase transitions. We believe that new and efficient representations will have to be developed which properly account for all the relevant degrees of freedom at hand. 



\begin{acknowledgement}
We acknowledge support by the Swiss National Science foundation (No.~PP00P2\_138932, 407540\_167186 NFP 75 Big Data, 200021\_175747, NCCR MARVEL).
\end{acknowledgement}

\bibliographystyle{spbasic}  
\bibliography{main} 

\begin{thebibliography}{38}
\providecommand{\natexlab}[1]{#1}
\providecommand{\url}[1]{{#1}}
\providecommand{\urlprefix}{URL }
\expandafter\ifx\csname urlstyle\endcsname\relax
  \providecommand{\doi}[1]{DOI~\discretionary{}{}{}#1}\else
  \providecommand{\doi}{DOI~\discretionary{}{}{}\begingroup
  \urlstyle{rm}\Url}\fi
\providecommand{\eprint}[2][]{\url{#2}}

\bibitem[{Axilrod and Teller(1943)}]{atm}
Axilrod BM, Teller E (1943) Interaction of the van der waals type between three
  atoms. J Chem Phys 11(6):299--300, \doi{10.1063/1.1723844},
  \urlprefix\url{http://scitation.aip.org/content/aip/journal/jcp/11/6/10.1063/1.1723844}

\bibitem[{Bader(1990)}]{AIM}
Bader RF (1990) Atoms in molecules. Wiley Online Library

\bibitem[{Bart\'ok et~al(2010)Bart\'ok, Payne, Kondor, and Cs\'anyi}]{GAP}
Bart\'ok AP, Payne MC, Kondor R, Cs\'anyi G (2010) Gaussian approximation
  potentials: The accuracy of quantum mechanics, without the electrons. Phys
  Rev Lett 104:136,403, \doi{10.1103/PhysRevLett.104.136403},
  \urlprefix\url{http://link.aps.org/doi/10.1103/PhysRevLett.104.136403}

\bibitem[{Bart\'ok et~al(2013)Bart\'ok, Kondor, and Cs\'anyi}]{soap_2013}
Bart\'ok AP, Kondor R, Cs\'anyi G (2013) On representing chemical environments.
  Phys Rev B 87:184,115, \doi{10.1103/PhysRevB.87.184115},
  \urlprefix\url{http://link.aps.org/doi/10.1103/PhysRevB.87.184115}

\bibitem[{Bart{\'o}k et~al(2017)Bart{\'o}k, De, Poelking, Bernstein, Kermode,
  Cs{\'a}nyi, and Ceriotti}]{CeriottiScienceUnified2017}
Bart{\'o}k AP, De S, Poelking C, Bernstein N, Kermode JR, Cs{\'a}nyi G,
  Ceriotti M (2017) Machine learning unifies the modeling of materials and
  molecules. Science Advances 3(12), \doi{10.1126/sciadv.1701816},
  \urlprefix\url{http://advances.sciencemag.org/content/3/12/e1701816},
  \eprint{http://advances.sciencemag.org/content/3/12/e1701816.full.pdf}

\bibitem[{Browning et~al(2017)Browning, Ramakrishnan, von Lilienfeld, and
  Roethlisberger}]{geneticAlgorithm_trainingSet}
Browning NJ, Ramakrishnan R, von Lilienfeld OA, Roethlisberger U (2017) Genetic
  optimization of training sets for improved machine learning models of
  molecular properties. J Phys Chem Lett 8(7):1351,
  \doi{10.1021/acs.jpclett.7b00038},
  \urlprefix\url{http://dx.doi.org/10.1021/acs.jpclett.7b00038}

\bibitem[{Christensen et~al(2017)Christensen, Faber, Huang, Bratholm,
  Tkatchenko, M\"uller, and von Lilienfeld}]{QMLcode1}
Christensen AS, Faber FA, Huang B, Bratholm LA, Tkatchenko A, M\"uller KR, von
  Lilienfeld OA (2017) Qml: A python toolkit for quantum machine learning
  \doi{10.5281/Zeno.817331}, https://github.com/qmlcode/qml

\bibitem[{Cs{\'a}nyi et~al(2004)Cs{\'a}nyi, Albaret, Payne, and
  Vita}]{lotf2004}
Cs{\'a}nyi G, Albaret T, Payne MC, Vita AD (2004) {``Learn on the Fly'': A}
  hybrid classical and quantum-mechanical molecular dynamics simulation. Phys
  Rev Lett 93:175,503

\bibitem[{De et~al(2016)De, Bartok, Csanyi, and Ceriotti}]{Sandip2016}
De S, Bartok AP, Csanyi G, Ceriotti M (2016) Comparing molecules and solids
  across structural and alchemical space. Phys Chem Chem Phys
  18:13,754--13,769, \doi{10.1039/C6CP00415F},
  \urlprefix\url{http://dx.doi.org/10.1039/C6CP00415F}

\bibitem[{Faber et~al(2017{\natexlab{a}})Faber, Christensen, Huang, and von
  Lilienfeld}]{faber2017alchemical}
Faber FA, Christensen AS, Huang B, von Lilienfeld OA (2017{\natexlab{a}})
  Alchemical and structural distribution based representation for improved qml.
  arXiv preprint arXiv:171208417

\bibitem[{Faber et~al(2017{\natexlab{b}})Faber, Hutchison, Huang, Gilmer,
  Schoenholz, Dahl, Vinyals, Kearnes, Riley, and von
  Lilienfeld}]{googlePaper2017}
Faber FA, Hutchison L, Huang B, Gilmer J, Schoenholz SS, Dahl GE, Vinyals O,
  Kearnes S, Riley PF, von Lilienfeld OA (2017{\natexlab{b}}) Fast machine
  learning models of electronic and energetic properties consistently reach
  approximation errors better than dft accuracy. https://arxivorg/abs/170205532
  \urlprefix\url{https://arxiv.org/abs/1702.05532}

\bibitem[{Fasshauer and McCourt(2016)}]{fasshauer2016kernel}
Fasshauer G, McCourt M (2016) Kernel-based approximation methods using Matlab.
  World Scientific

\bibitem[{Fias et~al(2017{\natexlab{a}})Fias, Heidar-Zadeh, Geerlings, and
  Ayers}]{StijnPNAS2017}
Fias S, Heidar-Zadeh F, Geerlings P, Ayers PW (2017{\natexlab{a}}) Chemical
  transferability of functional groups follows from the nearsightedness of
  electronic matter. Proceedings of the National Academy of Sciences
  114(44):11,633--11,638

\bibitem[{Fias et~al(2017{\natexlab{b}})Fias, Heidar-Zadeh, Geerlings, and
  Ayers}]{stjin2017}
Fias S, Heidar-Zadeh F, Geerlings P, Ayers PW (2017{\natexlab{b}}) Chemical
  transferability of functional groups follows from the nearsightedness of
  electronic matter. Proceedings of the National Academy of Sciences
  114(44):11,633--11,638

\bibitem[{Ghiringhelli et~al(2015)Ghiringhelli, Vybiral, Levchenko, Draxl, and
  Scheffler}]{descriptor_design_LASSO}
Ghiringhelli LM, Vybiral J, Levchenko SV, Draxl C, Scheffler M (2015) Big data
  of materials science: Critical role of the descriptor. Phys Rev Lett
  114:105,503, \doi{10.1103/PhysRevLett.114.105503},
  \urlprefix\url{http://link.aps.org/doi/10.1103/PhysRevLett.114.105503}

\bibitem[{Greeley et~al(2006)Greeley, Jaramillo, Bonde, Chorkendorff, and
  N{\o}rskov}]{greeley2006}
Greeley J, Jaramillo TF, Bonde J, Chorkendorff I, N{\o}rskov JK (2006)
  Computational high-throughput screening of electrocatalytic materials for
  hydrogen evolution. Nature materials 5(11):909--913

\bibitem[{Hansen et~al(2015)Hansen, Biegler, von Lilienfeld, M\"uller, and
  Tkatchenko}]{BoB}
Hansen K, Biegler F, von Lilienfeld OA, M\"uller KR, Tkatchenko A (2015)
  Interaction potentials in molecules and non-local information in chemical
  space. J Phys Chem Lett 6:2326, \doi{10.1021/acs.jpclett.5b00831},
  \urlprefix\url{http://dx.doi.org/10.1021/acs.jpclett.5b00831}

\bibitem[{Huang and von Lilienfeld(2016)}]{baml}
Huang B, von Lilienfeld OA (2016) Communication: Understanding molecular
  representations in machine learning: The role of uniqueness and target
  similarity. J Chem Phys 145(16):161102, \doi{10.1063/1.4964627},
  \urlprefix\url{http://dx.doi.org/10.1063/1.49646277}

\bibitem[{Huang and von Lilienfeld(2017)}]{huang2018}
Huang B, von Lilienfeld OA (2017) Chemical space exploration with molecular
  genes and machine learning. arXiv preprint arXiv:170704146

\bibitem[{Kennedy and O'Hagan(2000)}]{kennedy2000}
Kennedy MC, O'Hagan A (2000) Predicting the output from a complex computer code
  when fast approximations are available. Biometrika 87(1):1--13

\bibitem[{Kirkpatrick and Ellis(2004)}]{CCS}
Kirkpatrick P, Ellis C (2004) Chemical space. Nature 432(7019):823--823,
  \doi{10.1038/432823a}, \urlprefix\url{http://dx.doi.org/10.1038/432823a}

\bibitem[{Kitaura et~al(1999)Kitaura, Ikeo, Asada, Nakano, and Uebayasi}]{FMO}
Kitaura K, Ikeo E, Asada T, Nakano T, Uebayasi M (1999) Fragment molecular
  orbital method: an approximate computational method for large molecules. Chem
  Phys Lett 313(3-4):701--706, \doi{10.1016/S0009-2614(99)00874-X}

\bibitem[{von Lilienfeld(2018)}]{QMLessayAnatole}
von Lilienfeld OA (2018) Quantum machine learning in chemical compound space.
  Angewandte Chemie International Edition
  Http://dx.doi.org/10.1002/anie.201709686

\bibitem[{von Lilienfeld et~al(2015)von Lilienfeld, Ramakrishnan, Rupp, and
  Knoll}]{OAvL_FRD}
von Lilienfeld OA, Ramakrishnan R, Rupp M, Knoll A (2015) Fourier series of
  atomic radial distribution functions: A molecular fingerprint for machine
  learning models of quantum chemical properties. Int J Quantum Chem
  115(16):1084--1093, \doi{10.1002/qua.24912},
  \urlprefix\url{http://dx.doi.org/10.1002/qua.24912}

\bibitem[{Muto(1943)}]{atm2}
Muto Y (1943) Force between nonpolar molecules. J Phys-Math Soc Japan
  17:629--631

\bibitem[{Pilania et~al(2017)Pilania, Gubernatis, and Lookman}]{pilania2017}
Pilania G, Gubernatis JE, Lookman T (2017) Multi-fidelity machine learning
  models for accurate bandgap predictions of solids. Computational Materials
  Science 129:156--163

\bibitem[{Podryabinkin and Shapeev(2017)}]{Shapeev2017}
Podryabinkin EV, Shapeev AV (2017) Active learning of linearly parametrized
  interatomic potentials. Computational Materials Science 140:171--180

\bibitem[{Prodan and Kohn(2005)}]{nearsightedness}
Prodan E, Kohn W (2005) Nearsightedness of electronic matter. Proc Natl Acad
  Sci USA 102(33):11,635--11,638, \doi{10.1073/pnas.0505436102},
  \urlprefix\url{http://dx.doi.org/10.1073/pnas.0505436102}

\bibitem[{Ramakrishnan and von Lilienfeld(2015)}]{one_kernel_Raghu}
Ramakrishnan R, von Lilienfeld OA (2015) Many molecular properties from one
  kernel in chemical space. CHIMIA 69(4):182, \doi{10.2533/chimia.2015.182},
  \urlprefix\url{http://www.ingentaconnect.com/content/scs/chimia/2015/00000069/00000004/art00005}

\bibitem[{Ramakrishnan et~al(2014)Ramakrishnan, Dral, Rupp, and von
  Lilienfeld}]{gdb9_data}
Ramakrishnan R, Dral P, Rupp M, von Lilienfeld OA (2014) Quantum chemistry
  structures and properties of 134 kilo molecules. Sci Data 1:140,022,
  \doi{10.1038/sdata.2014.22},
  \urlprefix\url{http://dx.doi.org/10.1038/sdata.2014.22}

\bibitem[{Ramakrishnan et~al(2015{\natexlab{a}})Ramakrishnan, Dral, Rupp, and
  von Lilienfeld}]{DeltaPaper2015}
Ramakrishnan R, Dral P, Rupp M, von Lilienfeld OA (2015{\natexlab{a}}) {Big
  Data meets Quantum Chemistry Approximations: The $\Delta$-Machine Learning
  Approach}. J Chem Theory Comput 11:2087--2096,
  \doi{10.1021/acs.jctc.5b00099},
  \urlprefix\url{http://dx.doi.org/10.1021/acs.jctc.5b00099}

\bibitem[{Ramakrishnan et~al(2015{\natexlab{b}})Ramakrishnan, Hartmann,
  Tapavicza, and von Lilienfeld}]{ML-TDDFTEnrico2015}
Ramakrishnan R, Hartmann M, Tapavicza E, von Lilienfeld OA (2015{\natexlab{b}})
  Electronic spectra from {TDDFT} and machine learning in chemical space. J
  Chem Phys 143:084,111, {\tt http://arxiv.org/abs/1504.01966}

\bibitem[{Rappe et~al(1992)Rappe, Casewit, Colwell, III, and Skiff}]{uff}
Rappe AK, Casewit CJ, Colwell KS, III WAG, Skiff WM (1992) Uff, a full periodic
  table force field for molecular mechanics and molecular dynamics simulations.
  J Am Chem Soc 114(25):10,024--10,035, \doi{10.1021/ja00051a040},
  \urlprefix\url{http://dx.doi.org/10.1021/ja00051a040}

\bibitem[{Rasmussen and Williams(2006)}]{GP}
Rasmussen C, Williams C (2006) Gaussian Processes for Machine Learning.
  Adaptative computation and machine learning series, University Press Group
  Limited, \urlprefix\url{https://books.google.ch/books?id=vWtwQgAACAAJ}

\bibitem[{Ruddigkeit et~al(2012)Ruddigkeit, van Deursen, Blum, and
  Reymond}]{gdb17}
Ruddigkeit L, van Deursen R, Blum LC, Reymond JL (2012) Enumeration of 166
  billion organic small molecules in the chemical universe database gdb-17. J
  Chem Inf Model 52(11):2864--2875, \doi{10.1021/ci300415d},
  \urlprefix\url{http://dx.doi.org/10.1021/ci300415d}

\bibitem[{Rupp et~al(2012)Rupp, Tkatchenko, M\"uller, and von Lilienfeld}]{CM}
Rupp M, Tkatchenko A, M\"uller KR, von Lilienfeld OA (2012) Fast and accurate
  modeling of molecular atomization energies with machine learning. Phys Rev
  Lett 108(5):058,301, \doi{10.1103/PhysRevLett.108.058301},
  \urlprefix\url{http://dx.doi.org/10.1103/PhysRevLett.108.058301}

\bibitem[{Samuel(2000)}]{WhatIsML}
Samuel AL (2000) Some studies in machine learning using the game of checkers.
  IBM Journal of research and development 44(1.2):206--226

\bibitem[{Todeschini and Consonni(2008)}]{todeschini2008}
Todeschini R, Consonni V (2008) Handbook of molecular descriptors, vol~11. John
  Wiley \& Sons

\end{thebibliography}

\end{document}